\documentclass[twocolumn,showpacs,preprintnumbers,superscriptaddress,amsmath,amssymb]{revtex4}
\usepackage{dcolumn}% Align table columns on decimal point
\usepackage{bm}% bold math
\usepackage{amsmath}
\usepackage{multirow}
\usepackage{graphicx}

\begin{document}

 %\twocolumn
 %\baselineskip = 1.5\baselineskip

 \newcommand{\re}{\mathop{\mathrm{Re}}}
 \newcommand{\im}{\mathop{\mathrm{Im}}}
 \newcommand{\D}{\mathop{\mathrm{d}}}
 \newcommand{\I}{\mathop{\mathrm{i}}}
 \newcommand{\E}{\mathop{\mathrm{e}}}
 \newcommand{\unite}[2]{\mbox{$#1\,{\rm #2}$}}
 \newcommand{\myvec}[1]{\mbox{$\overrightarrow{#1}$}}
 \newcommand{\mynor}[1]{\mbox{$\widehat{#1}$}}
 \newcommand{\rmsemit}{\mbox{$\tilde{\varepsilon}$}}
 \newcommand{\mean}[1]{\mbox{$\langle{#1}\rangle$}}

\title{Generation of Relativistic Electron Bunches with Arbitrary Current Distribution \\
via Transverse-to-Longitudinal Phase Space Exchange}% Force line breaks with \\
\author{P. Piot} \affiliation{Northern Illinois Center for
Accelerator \& Detector Development and Department of Physics,
Northern Illinois University, DeKalb IL 60115,
USA}\affiliation{Accelerator Physics Center, Fermi National
Accelerator Laboratory, Batavia, IL 60510, USA}
\author{Y.-E Sun}
\affiliation{Accelerator Physics Center, Fermi National Accelerator
Laboratory, Batavia, IL 60510, USA}
\author{J. G. Power}\affiliation{High Energy Physics Division, Argonne
National Laboratory, Argonne IL 60439, USA}
\author{M. Rihaoui} \affiliation{Northern Illinois Center for
Accelerator \& Detector Development and Department of Physics,
Northern Illinois University, DeKalb IL 60115,
USA}
\preprint{FERMILAB-PUB-09-265-APC}
\date{\today}% It is always \today, today,
             %  but any date may be explicitly specified

\begin{abstract}
We propose a general method for tailoring the current distribution of relativistic electron bunches. The technique relies on a recently proposed method to exchange the longitudinal phase space emittance with one of the transverse emittances. The method consists of transversely shaping the bunch and then converting its transverse profile into a current profile via a transverse-to-longitudinal phase-space-exchange beamline.  We show that it is possible to tailor the current profile to follow, in principle, any desired distributions.  We demonstrate, via computer simulations, the application of the method to generate trains of microbunches with tunable spacing and  linearly-ramped current profiles. We also briefly explore potential applications of the technique. 
\end{abstract}
% PACS, the Physics and Astronomy
                             % Classification Scheme.
%\keywords{Suggested keywords}%Use showkeys class option if keyword
                              %display desired%
\maketitle
\section{Introduction}
Modern applications of particle accelerators generally rely on phase space manipulations within one or two degrees of freedom. These manipulations are necessary to tailor the phase space distributions to match the requirements of the front-end
applications. Usually,  phase space matching is done in a root-mean-square (rms) sense. However, there has been an increasing demand for more precise phase space control. In particular, electron bunches with tailored temporal distributions are often desired. For instance, in accelerator-driven short-wavelength light sources based on the free-electron laser (FEL) principle and using the so-called high gain harmonic generation (HGHG), a uniform current density is needed~\cite{hghg}. On the other hand the
production of trains of microbunches with a given bunch-to-bunch separation is foreseen to pave the road toward compact light source operating in the super-radiant regime  at wavelength comparable or larger than the typical microbunch length~\cite{gover}.
Finally, novel accelerator concepts based on beam-driven acceleration mechanisms, e.g. plasma or dielectric wakefield-based acceleration~\cite{petra,wei,travish,pwfa}, would greatly benefit from linearly ramped or trapezoidal current profiles~\cite{bane,nlpwfa} to significantly increase the transformer ratio -- the energy gain of the accelerated bunch over the energy loss of the driving bunch~\cite{petra}. Furthermore the performance of the aforementioned acceleration schemes,  could be significantly improved if driven by a train of microbunches to resonantly drive wakefields~\cite{muggliprstab} or train of linearly ramped microbunches~\cite{power,jing} to improve the transformer ratio.

To address these needs, several techniques aimed at tailoring the current profile of electron bunches have been explored. Linearly-ramped current profiles can be produced by imparting nonlinear distortions in the longitudinal phase space~\cite{pioterl,england1,piotaac08}. A successful experimental implementation of a nonlinear optical lattice was reported in Ref.~\cite{england}. The generation of a train of microbunches by shaping the temporal profile associated to a photocathode drive laser in a photoemission electron source were explored via numerical simulations~\cite{bosco,ychuang,yuelin}. An alternative method using an interceptive mask located in a dispersive section, first  proposed in Reference~\cite{nguyen}, was experimentally demonstrated~\cite{muggli}. Tailoring of the current profile to enhance the operation of a FEL operating in the HGHG regime via an ad hoc temporal shaping of the drive-laser was discussed in Reference~\cite{massimo}.

Each of these methods has limitations. The techniques based on shaping of the photocathode drive laser distribution are prone to the space-charge effects which are prominent at low energies and tend to smear the impressed shaping. Although it has been pointed out that the density modulation is indeed converted into an energy modulation and can be later recovered by a proper longitudinal phase space  manipulation, the final  bunching factor is significantly lower~\cite{neumann}. Similarly the techniques based on nonlinear transformations generally have a limited tunability.

In this paper we propose an alternative method to tailor the current distribution of relativistic  electron bunches extending on our earlier work~\cite{piotaac082,yine2}. The technique relies on the transverse-to-longitudinal phase  space exchange technique.  After discussing the theoretical background for our proposal in Section~\ref{sec:intro}, we demonstrate the generation of various current profiles via computer  modeling in Section~\ref{sec:demo}. We finally discuss  a possible implementation and its applications in Section~\ref{sec:appl1}. 

\section{Theoretical background\label{sec:intro}}
\begin{figure}[hbp]
\includegraphics[angle=0,width=0.5\textwidth]{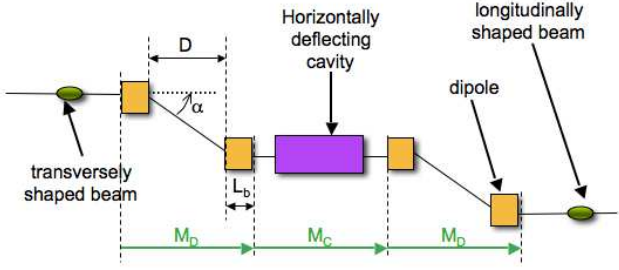}
\caption{(Color) Overview of the proposed technique to produce
relativistic electron bunch with arbitrary current profile. }
\label{fig:scheme}
\end{figure}

The proposed method for producing beams with arbitrary current profile is illustrated in Figure~\ref{fig:scheme}. A 
transversely-shaped electron bunch with the transverse density distribution $\Phi_{\perp}(x,y)$ is manipulated through a 
transverse-to-longitudinal emittance exchange (EEX) beamline to convert the corresponding horizontal profile $P(x)=\int dy \Phi_{\perp}(x,y)$ into the longitudinal (temporal) coordinate; see Fig.~\ref{fig:scheme}.

The backbone of our proposed scheme is the transverse-to-longitudinal phase space exchange. The emittance exchange was first proposed in the context of B-factory~\cite{orlov} as a way to achieve very small $\beta^*$ values at the interaction point. The scheme was later explored as a possible alternative for mitigating microbunching instability in bright electron beams~\cite{emma} or for improving the performance of single-pass FELs~\cite{emma2}. There are several solutions capable of performing this phase space exchange~\cite{fliller,piotUCLA}. The simplest solution devised to date consists of a horizontally-deflecting cavity, operating on the TM$_{110}$ mode, flanked by two identical horizontally-dispersive sections henceforth referred to as ``dogleg"~\cite{kim}; see Fig.~\ref{fig:scheme}. An experiment conducted at the Fermilab's A0 photoinjector facility recently demonstrated the exchange of transverse and longitudinal emittances~\cite{amber}.

We first elaborate on the emittance exchange principle and analyze the single-particle dynamics in the four dimensional phase space $(x,x',z,\delta)$, an initial electron with coordinate $\underline{\bf \widetilde{X}_0 }\equiv(x_0,x_0',z_0,\delta_0)$ is mapped to its final coordinate $ \underline{\bf \widetilde{X}}\equiv(x,x',z,\delta)$ accordingly to $\underline{\bf{X}}=R \underline{\bf{X}_0}$ wherein $R$ is the $4\times4$ transport matrix; the $\widetilde{~}$ stands for the transposition operator. In our notation underlined vectors belong to the four-dimensional trace space $(x,x',z,\delta)$. The matrix of a dogleg composed of two rectangular bends of length $L_b$ and bending angle $\alpha$ separated by a drift space $D$ has the form
\begin{eqnarray}
M_{D}&= & \left[ \begin {array}{cccc}
1&L&0&\eta\\\noalign{\medskip}0&1&0&0\\\noalign{\medskip}0&\eta&1&\xi\\\noalign{\medskip}0&0&0&1\end
{array} \right]
\end{eqnarray}
where the matrix elements are related to the machine parameters by the following: $L= {\frac {2\,c_{\alpha} {\it L_b}+D}{c_{\alpha}^{2}}}$, $\xi= {\frac {2c_{\alpha} s_{\alpha} {\it L_b}+s_{\alpha}^3  D-2\,{\it L_b}\alpha c_{\alpha}^{2}}{s_{\alpha}c_{\alpha}^{2}}}$ and $\eta={\frac {D+2c_{\alpha} {\it L_b}-c_{\alpha} ^{2}({2\it
L_b+D})}{s_{\alpha } c_{\alpha} ^{2}}}$, with the short-handed notation $(c,s)_{\alpha}= (\cos, \sin) \alpha$~\cite{yine}. The transfer matrix associated to a deflecting cavity of length $L_c$ and deflecting strength $\kappa$ is~\cite{emma,edwardsNote}:
\begin{eqnarray} \label{eq:cavdon}
M_{C}=\left[ \begin {array}{cccc} 1&{\it L_c}&\kappa{\it
L_c}/2&0\\\noalign{\medskip}0&1&\kappa&0\\\noalign{\medskip}0&0&1&0\\\noalign{\medskip}\kappa&\kappa{\it
L_c}/2&{\kappa}^{2}{\it L_c}/4&1\end {array} \right].
\end{eqnarray}
A complete emittance exchange requires $\kappa=-1/\eta$~\cite{kim} and the total matrix becomes:
\begin{eqnarray}
M=M_{D} M_{C} M_{D} = \left[\begin{array}{cc} A &B \\ C &D
\end{array} \right],
\end{eqnarray}
where the elements of matrix $A$ are all zero except $A_{1,2}=\frac{
L_c}{4}$, $B=-\frac{1}{\eta}\left[ \begin {array}{cc} { ({{\it
L_c}+4L})/4}&{ ({4\xi L-4{\eta}^{2}+\xi{\it
L_c}})/4} \\
1&{ {\xi}}
\end{array} \right]$, $C$ has the same form as $B$ but its diagonal elements are
reversed, and $D=\frac{\it L_c}{4{\eta}^{2}}\left[
\begin {array}{cc} {\xi}&{{\xi}^{2}} \\
1&{\xi}
\end{array} \right]$. For a thin cavity ($L_c=0$), the matrices $A$ and $D$ vanish and the exchange is perfect~\cite{kim}. In this latter case we have a direct mapping of the transverse to longitudinal coordinates and vice versa. In particular we have:
\begin{eqnarray} \label{eq:thinlenstransform}
\left\{ \begin {array}{ll} z =& -\frac{\xi}{\eta}x_0-\frac{L\xi-\eta^2}{\eta}x_0'\\
\delta =& -\frac{1}{\eta}x_0-\frac{L}{\eta}x_0'
\end{array}\right.
\end{eqnarray}

In general, when including thick lens effects,  the final beam matrix is $\underline {\Sigma} \equiv \mean{\underline {{\bf X}}\underline {{\bf \widetilde{X}}}}= \underline {M} \underline {\Sigma_0} \underline {\widetilde{M}}$ is not block anti-diagonal and the final geometric emittances are given to first order by~\cite{emma}
\begin{eqnarray} \label{eq:emitgrowth}
\left\{ \begin {array}{ll} \varepsilon_x^2 &=  \varepsilon_{z,0}^2 \left( 1 + \varrho_0 \lambda^2  \right) \\ 
\varepsilon_z^2 &=  \varepsilon_{x,0}^2 \left( 1 + \frac{\lambda^2}{\varrho_0}  \right)  
\end{array}\right.  
\end{eqnarray}
where $\varrho_0 \equiv \varepsilon_{x,0}/\varepsilon_{z,0}$ is the initial transverse-to-longitudinal emittance ratio and $\lambda^2$ is given by~\cite{yine}
\begin{eqnarray}
{\lambda}^2=\frac{L_c^2(1+\alpha_{x,0}^2)[\xi^2+(\xi\alpha_{z,0}-2\beta_{z,0})^2]}{64\eta^2\beta_{x,0}\beta_{z,0}}.
\end{eqnarray}
Here $\alpha_{i,0}$ and $\beta_{i,0}$ are the Courant-Snyder (C-S) parameters associated to the horizontal ($i=x$) and longitudinal ($i=z$) degrees of freedom. The quantity
$\lambda^2$ can be minimized by a proper choice of either longitudinal or transverse C-S parameters. Here we opt for minimizing $\lambda$ with respect to the longitudinal C-S parameters. Since the bunch length (i.e. $\beta_{z,0}$) is fixed by the upstream beamline and electron source settings for most of the cases discussed in this paper, $\lambda$ is minimized with respect to $\alpha_{z,0}= -\frac{\langle z_0\delta_0\rangle}{\varepsilon_{z,0}}$, the
optimum value is $\hat{\alpha}_{z,0}=\,{\frac {{\it \beta_z}}{\xi}}$ corresponding to an incoming  longitudinal phase space chirp ${\cal C}\equiv\frac{d\delta}{dz}\big|_0=-\frac{1}{\xi}$ that produces a minimum bunch length at the cavity location. \\

In order to illustrate how this phase space exchange mechanism could be taken advantage of  to tailor the longitudinal distribution of an electron bunch, we consider a beam with an initial six-dimensional Gaussian trace space distribution
\begin{eqnarray}
\Phi_0({{\bf X}_0})=\frac{n_0}{(2\pi)^3\varepsilon_{x,0}
\varepsilon_{y,0} \varepsilon_{z,0}}\exp[-\frac{1}{2}{\widetilde{\bf
X}_0}\Sigma_0^{-1}{\bf X}_0]
\end{eqnarray}
where ${\widetilde{\bf X}_0} \equiv (x_0,x'_0,y_0,y'_0,z_0,\delta_0)$, and
$\Sigma \equiv \mean{\bf{X}_0\widetilde{\bf{X}}_0}$ is the
six-dimensional covariance matrix associated to the beam. We further assume the initial covariance matrix to be uncoupled, i.e. to have the form
$\Sigma_0=\mbox{diag}(\Sigma_{x,0},\Sigma_{y,0},\Sigma_{z,0})$ with
\begin{eqnarray}
\Sigma_{i,0} \equiv \varepsilon_{i,0}\left[
\begin {array}{cc}
  \beta_{i,0} & -\alpha_{i,0} \\\noalign{\medskip}
 -\alpha_{i,0} & \gamma_{i,0} \\\noalign{\medskip}
\end {array}
\right],
\end{eqnarray}
where the index $i=x,y,z$ stands for the considered degree of freedom and $\gamma_{i,0}\equiv (1+\alpha_{i,0}^2)/\beta_{i,0}$.

We suppose the initial distribution function can be transversely shaped with a ``transmission" function $T(x_0,y_0)$. The function $T(x_0,y_0)$ can model a mask [if so its value is either 1 (within the mask aperture) or 0 (elsewhere)] or be used to account for a transverse density that deviates from the Gaussian distribution.  Therefore the four-dimensional trace space distribution upstream of the EEX beamline is 
\begin{eqnarray}
\Phi_0^+({{\bf X}_0}) &= & \Phi_0^-({{\bf
X}_0};0) \int_{-\infty} ^{+\infty} \frac{dy_0}{\sqrt{2\pi}
\sigma_{y,0}}
\\ \nonumber & & \times  \exp \left(-\frac{y_0^2}{2\beta_{y,0}
\varepsilon_{y,0}} \right) T(x_0,y_0) \\ \nonumber & \equiv &
\Phi_0^-({{\bf X}_0};0)  \tau ({\bf{\bf X}_0}).
\end{eqnarray}
%%%%%%%%%%%%%%%%%%%%%
Downstream of the EEX beamline the final distribution is of the form
\begin{eqnarray}
\Phi({{\bf X}_0;s}) &= & \Phi(R^{-1}{{\bf X}_0};0) \tau(R^{-1}{{\bf X}_0}),
\end{eqnarray}
and the longitudinal trace space is given by
\begin{eqnarray}\label{eq:finalzd}
\Phi(z,\delta) &= & \frac{n_0}{\sqrt{2\pi}\varepsilon_{z,0}} \exp \left(-\frac{z^2}{2\beta_z\varepsilon_{x,0}} \right) \nonumber \\
&& \times \exp \left[-\frac{\beta_z}{2\epsilon_{x,0}} \left(\delta - \frac{\alpha_z}{\beta_z} z\right)^2\right] \nonumber \\
 && \times  \tau(N_{15}z + N_{16} \delta),
\end{eqnarray}%
where $N_{i,j}$ are the elements of the matrix $N \equiv R^{-1}$. The latter equation describes a two-dimensional Gaussian distribution in $(z,\delta)$ modulated by the function $\tau$.  Integration over $\delta$ provides the longitudinal profile downstream of the exchanger:
\begin{eqnarray}
P(z) & = & \frac{n_0}{2\pi \varepsilon_{x,0}} \exp
\left(-\frac{z^2}{2\beta_z\varepsilon_{x,0}} \right) \Theta(z),
\end{eqnarray}
with $\Theta(z)\equiv \int_{-\infty}^{+\infty} \exp \left[-\frac{\beta_z}{2\epsilon_{x,0}} \left(\delta - \frac{\alpha_z}{\beta_z} z\right)^2\right]  \tau(N_{15}z + N_{16} \delta) d\delta $. If the mask has no y-dependence [$y^+(x) \rightarrow + \infty$ and $y^-(x) \rightarrow - \infty$ ], $\tau$ is a function of $x$ only. \\

\begin{figure}[htp]
\includegraphics[angle=0,width=0.5\textwidth]{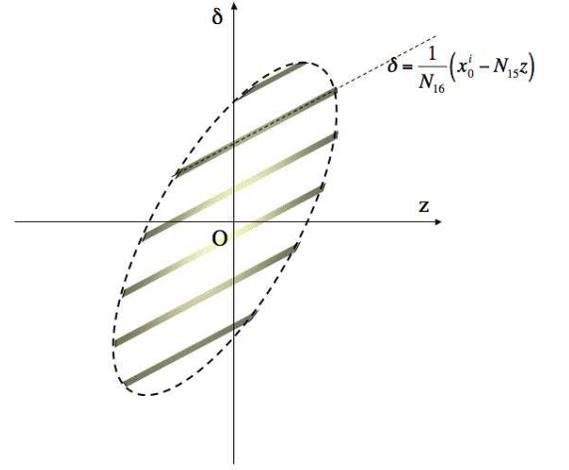}
\caption{Rendition of a stratified longitudinal phase space generated downstream of the EEX beamline with an incoming transversely-modulated beam. } \label{fig:longphsp}
\end{figure}

We first consider the case of $\tau(x_0) = \sum_{i=1}^N \delta(x-x_0^i)$ where $\delta(...)$ is the Dirac function.  This can be achieved, e.g.,  by intercepting the incoming beam with a series of vertical slits with center located at $x_0^i$. In such a case the longitudinal phase space is described by a two-dimensional Gaussian
distribution with C-S parameters $(\alpha_z,\beta_z)$, emittance $\varepsilon_{x0}$, which is non-vanishing only at locations described by the family of lines with equation $\delta(z)= (x_0^i+N_{15}z)/N_{16}$ in the longitudinal phase space; see Fig.~\ref{fig:longphsp}.
The associated projection along the longitudinal coordinate $z$ is
\begin{eqnarray}
P(z)&=&\frac{n_0}{2\pi\varepsilon_{z,0}} \exp \left( -\frac{z^2}{2\beta_z\varepsilon_{x,0}}\right) \sum_{i=1}^{M} \exp \left\{ -\frac{\beta_z}{2\varepsilon_{x,0} N_{16}^2} \right. \nonumber \\
& &\left.  \times \left[ x_0^i- z\left(N_{15} -\frac{\alpha_z}{\beta_z} N_{16} \right) \right]^2 \right\}, 
\end{eqnarray}
representing a series of Gaussian microbunches each of length $\sigma_{z,m}\equiv \sqrt{\varepsilon_{x,0} N_{16}^2/\beta_z}$. The microbunches are longitudinally separated when $4\sigma_{z,m} \lesssim x_0^i/[N_{15} -\alpha_z/\beta_z N_{16}]$. Even when the $P(z)$ projection is not strongly  modulated, the local correlations in the longitudinal trace space can be taken advantage of to enhance the modulation. This can be done by using a dispersive section downstream of the EEX beamline with its longitudinal dispersion $\xi'$ matched to the chirp of the beamlets, i.e. $\xi'=-N_{15}/N_{16}$. A simple estimate of $\xi'$ in the thin-lens approximation gives the required value to be
\begin{eqnarray}
\xi' = \xi-\frac{\eta^2}{L} >0.
\end{eqnarray}
In our convention [head of the bunch is at positive time (or $z$-coordinate)] the required value is of opposite sign of the value provided by, e.g., a  standard four-bend magnetic chicane. Such a longitudinal dispersion can be generated by a four-bend chicane with quadrupoles inserted between the two first and two last dipoles~\cite{bc}.\\

We now consider the more general case of a mask with upper ($y>0$) and lower ($y<0$)  boundaries respectively respectively described by the functions $y_0^+(x_0)$ and $y_0^-(x_0)$. In such case we have 
\begin{eqnarray}
\tau({\bf X}_0) &=& \tau(x_0) =\int_{y_0^-(x_0)} ^{y_0^+(x_0)} \frac{dy_0}{\sqrt{2\pi} \sigma_{y,0}}
\\ \nonumber & & \times  \exp \left(-\frac{y_0^2}{2\beta_{y,0}
\varepsilon_{y,0}} \right) T(x_0,y_0). 
\end{eqnarray}
Taking $y_0^-=-y_0^+\equiv \nu(x_0)$, where the function $\nu(x)$ describes the mask shape,  and $T(x_0)=1$ for $y_0(x_0)\in[y_0^-(x_0),y_0^+(x_0)]$  yields 

\begin{eqnarray}
\tau(x_0) &=& 2~\mbox{erf} \left[ \frac{\nu(x_0) }{\sqrt{2} \sigma_{y,0}}\right]. 
\end{eqnarray}

The general form of the longitudinal bunch profile downstream of the EEX beamline is 
\begin{eqnarray}
P(z) &=& \frac{n_0}{2\pi\varepsilon_{z,0}} \exp \left( -\frac{z^2}{2\beta_z\varepsilon_{x,0}}\right)  \int_{-\infty}^{+\infty} d\delta  \exp \\  \nonumber 
 & &  \left[-\frac{\beta_z}{2\epsilon_{x,0}} \left(\delta - \frac{\alpha_z}{\beta_z} z\right)^2\right]\mbox{erf}\left[\frac{\nu (N_{15}z + N_{16} \delta)}{\sqrt{2}\sigma_{y,0}}\right]. 
\end{eqnarray}
The latter equation provides some insight on the limitation of the discussed pulse shaping technique. For instance, the convolution-like integral indicates that current profiles with spectral components beyond the spatial frequency $f_x \sim 1/[2\pi\sqrt{\beta_z \varepsilon_{x,0}}]$ will be filtered out. Therefore the generation of pulses with sharp features would require the final longitudinal emittance and betatron function to be small.  
\section{Validation of the method via single-particle dynamics simulations\label{sec:demo}}
We first validate the technique proposed in the previous Section with the help of single-particle dynamics simulations. We consider an incoming beam with Gaussian distribution in the three degrees of freedom. The beam is intercepted by a mask with transverse transmission $T(x,y)$  and then passes through the emittance exchanger. The beamline geometric parameters are gathered in Table~\ref{tab:beamline} and correspond to the proof-of-principle experiment being considered at the Argonne Wakefield Accelerator (AWA) facility~\cite{awa}. The transversely-deflecting cavity consists of a 1/2-1-1/2 cell cavity optimized to minimize steering effects~\cite{shi}. 
\begin{table}[hbt]
\caption{Geometric parameters and properties associated to the EEX beamline considered in this paper; see also Fig.~\ref{fig:scheme}.\label{tab:beamline} }

\begin{center}
\begin{tabular}{l c c c}\hline\hline\
parameter & symbol & value & unit \\
\hline
Bending angle & $\alpha$  & 20 & deg   \\
Bend effective projected length & $L_b$  &0.308 & m \\
Projected distance between bends &  $D$    & 0.97 & m  \\
Dogleg final dispersion & $\eta $ & 0.51 & m\\
Dogleg longitudinal dispersion & $\xi$ &0.16 & m\\
\hline \hline
\end{tabular}
\end{center}
\end{table}
The numerical beam dynamics simulations presented in this section  were performed with the particle tracking code {\sc elegant}~\cite{elegant} and do not include any collective effects. The influence of collective effects is investigated in the next section when considering a realistic beamline configuration along with exploring practical applications of the pulse shaping technique. 

\begin{figure}
\includegraphics[scale = 0.70]{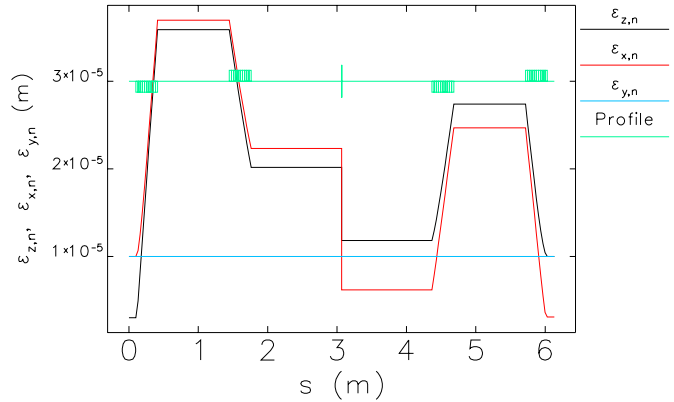}
\caption{Evolution of the transverse horizontal (black), vertical (cyan) and longitudinal (red) emittances along the EEX beamline with parameters given in Table.~\ref{tab:beamline}. The green symbols indicate the axial locations of the four dipoles (shown as rectangles) and deflecting cavity (vertical line at $s\simeq 3$~m). The initial normalized emittance partition is $(\varepsilon^n_{x,0},\varepsilon^n_{y,0},\varepsilon^n_{z,0})=(1,1, 10)$~$\mu$m. 
\label{fig:emitEXrms}}
\end{figure}

In {\sc elegant} the deflecting cavity was modeled by its thick-lens transfer matrix which was found to be in excellent agreement with the transfer matrix numerically devised by tracking macroparticles in the 3D electromagnetic field map computed with the {\sc microwave studio\textregistered}  eigensolver~\cite{CSTMW}; see Appendix A. An example of emittance evolution along the emittance exchanger with parameters tabulated in Table.~\ref{tab:beamline}, is shown in Fig.~\ref{fig:emitEXrms}.

\subsection{Limiting effects in an emittance-exchange beamline \label{sec:dilutingeffects}}
\begin{figure}[hhhhhhhhh!!!!!!!!!!!!!!]
\includegraphics[scale = 0.45]{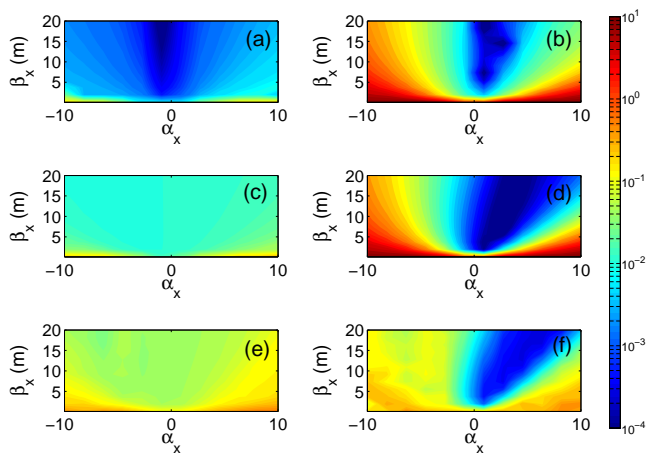}
\caption{Fractional horizontal (left column) and longitudinal (right column) emittance dilutions at the exit of the exchanger beamline versus the incoming horizontal Courant-Snyder beam parameters. These quantities are respectively defined as $\delta \varepsilon_x \equiv \varepsilon_x/\varepsilon_{z,0}-1$ and  $\delta \varepsilon_z \equiv \varepsilon_z/\varepsilon_{x,0}-1$. The rows correspond to different incoming normalized emittance partitions from top to bottom $(\varepsilon^n_{x,0},\varepsilon^n_{z,0})$=(10,1), (1,1) and (1,10) [all numbers are in micrometers]. The longitudinal phase space parameters used in these calculations are a bunch length $\sigma_{z,0}=400$~$\mu$m, and a chirp ${\cal C}=6$~m$^{-1}$.  
\label{fig:dilution}}
\end{figure}

Before proceeding with validating the pulse shaping method described in the previous section, we explore the domain of perfect phase space exchange. Several sources can prevent a perfect emittance exchange. One of them, described in  the previous section, is due to coupling introduced by the thick lens model of the deflecting mode cavity. Although minimizing this emittance dilution requires the chirp to be tuned to minimize the bunch length at the cavity location, the emittance is still diluted and the dilution term can be further minimized by selecting proper incoming  horizontal C-S parameters as illustrated in Fig.~\ref{fig:dilution}. In the latter Figure we introduced the normalized emittances defined for relativistic beams as $\varepsilon^n_{i,0}\equiv \gamma\varepsilon_{i,0}$ where $i=x,y,z$. 

\subsection{Generation of stratified longitudinal phase spaces and train of microbunches}
\begin{figure}[hhhhhhhhh!!!!!!!!!!!!!!]
\includegraphics[scale = 0.48]{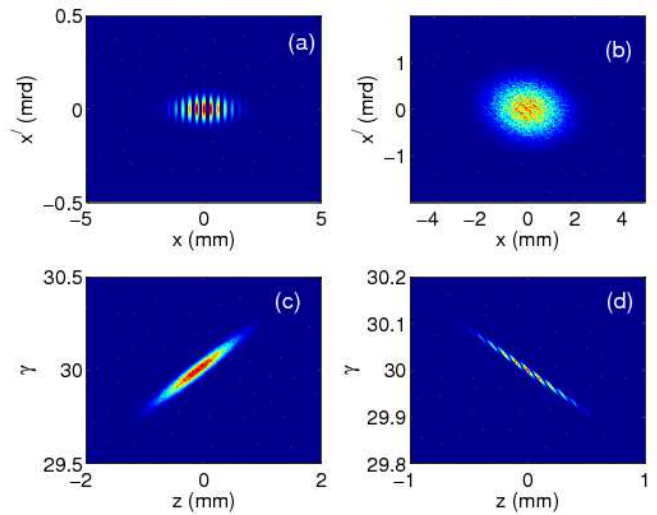}
\caption{Example of conversion of a transverse modulation (a) into a modulation on the longitudinal phase space (d).  The plots (a) and (c) show the initial horizontal and longitudinal phase space while the plots (b) and (d) show the corresponding final phase spaces downstream of the EEX beamline. The initial emittance partition is $(\varepsilon^n_{x,0},\varepsilon^n_{z,0})$=(1,10)~$\mu$m. 
\label{fig:exampleshowerimpact}}
\end{figure}

We first demonstrate the generation of a train of microbunches using an interceptive mask located upstream of emittance exchanger beamline. For these simulations the incoming Gaussian beam is intercepted by a 3~mm thick tungsten mask. The slits width is taken to be 100~$\mu$m and the edge-to-edge separation between the slits, $\delta w$, is variable. The simulations of the beam interaction with the mask, including scattering effects, is executed with {\sc shower}~\cite{shower} and the beam distribution downstream of the mask is then used in {\sc elegant} to model the emittance exchanger beamline; see Fig.~\ref{fig:exampleshowerimpact} \\

\begin{figure}[hhhhhhhhh!!!!!!!!!!!!!!]
\includegraphics[scale = 0.48]{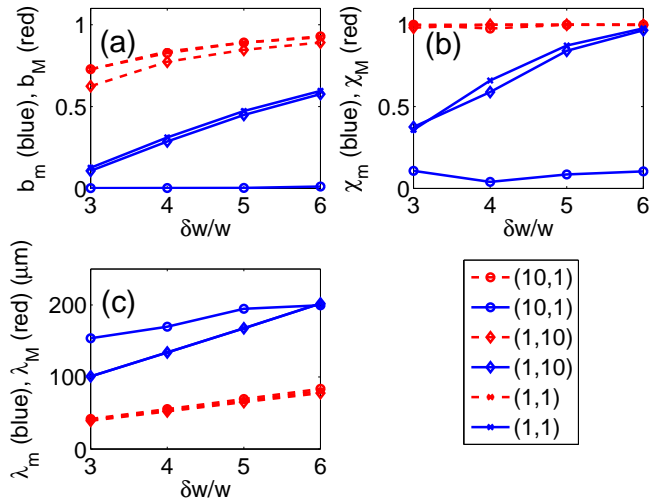}
\caption{Bunching factor (a), contrast ratio (b) and modulation wavelength (c) obtained for a longitudinally modulated beam produced via interception with thick mask upstream of the EEX beamline. The slits width is $w=100$~$\mu$m and the slits separation $\delta w$ is variable. The parameters with subscript $_m$ (and shown as blue traces) are computed using the longitudinal beam distribution downstream of the EEX beamline while the parameters with subscript $_M$ (and shown as red traces) are obtained after the removal of the chirp associated to the beamlet (in this latter case, the beamlets are made upright) . The parameters are calculated for various initial normalized emittance partitions ($\varepsilon^n_{x,0},\varepsilon^n_{z,0}$) shown in the legend. The beam Lorentz factor for these simulations is $\gamma=30$.  
\label{fig:param}}
\end{figure}

A series of simulations were performed for three cases of incoming transverse-to-longitudinal emittance ratio $\rho_0$. The horizontal C-S parameters were chosen for an optimum emittance exchange in accordance with Fig.~\ref{fig:dilution} and the beam's Lorentz is $\gamma=30$. The obtained current profiles were characterized by  computing the modulation wavelength $\lambda_m$, and the associated bunching factor $b_m\equiv b(k_m)$ where 
\begin{eqnarray}
b(k)=\int_{-\infty}^{+\infty} P(z) e^{ikz} dz
\end{eqnarray}
and $k_m=2\pi/\lambda_m$.  The contrast ratio
\begin{eqnarray}
\chi=\frac{\hat{P} - \check{P}}{\hat{P} + \check{P}}, 
\end{eqnarray}
where $\hat{P} \equiv \mbox{max} [P(z)]$ and $\check{P} \equiv \mbox{min} [P(z)]$, was also calculated. The resulting values for $\lambda_m$, $b_m$ and the corresponding contrast ratio $\chi_m$  are presented in Figure~\ref{fig:param}. We also include the parameters computed after removal of the local correlations such to result in upright longitudinal trace space for the beamlets: the modulation wavelength $\lambda_M$, associated bunching factor $b_M\equiv {b'}(k_M)$ (where $k_M\equiv 2\pi/\lambda_M$ and $b'$ is the bunching factor computed after the longitudinal phase space manipulation), and the contrast ratio $\chi_M$. When the final horizontal and longitudinal emittances are equal [$(\varepsilon^n_{x,0},\varepsilon^n_{z,0})$=(1,10)~$\mu$m] or when the final longitudinal emittance is smaller than the horizontal one [$(\varepsilon^n_{x,0},\varepsilon^n_{z,0})$=(1,10)~$\mu$m], a clear bunching is observed (for most of the cases $b_m>0.1$ and $\chi_m \gtrsim 0.5$). For the case when the final longitudinal emittance is larger [$(\varepsilon^n_{x,0},\varepsilon^n_{z,0})$=(10,1)~$\mu$m],  the bunching factor is small, i.e. $b_m \lesssim 10^{-2}$ but the longitudinal trace space is clearly stratified and the removal of the local correlations leads to strong bunching: the bunching factor becomes $b_M \gtrsim 0.5$ and the contrast ratios for all cases  are close to unity.  The modulation wavelength for the considered cases was $0.1 \le \lambda_m \le 0.2$~mm and $0.035 \lambda_M \le 0.080$~mm.

\begin{figure}[hhhhhhhhh!!!!!!!!!!!!!!]
\includegraphics[scale = 0.38]{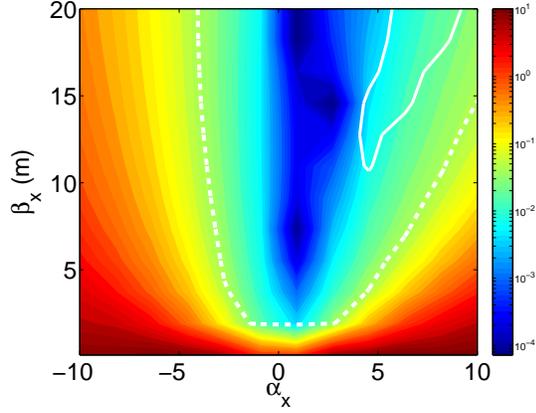}
\caption{Locus of C-S parameters that provide an upright longitudinal trace space for the beamlets (white solid line). The false color map represents the emittance dilution $\delta\epsilon_z$ as a function of initial C-S parameters and the dashed white line indicate the limit where emittance dilution is 5\%. The initial normalized emittance partition is $(\varepsilon^n_{x,0},\varepsilon^n_{z,0})=(10,1)$~$\mu$m. The solid white line is computed numerically and corresponds to the Eq.~\ref{eqn:betaoptimi}. 
\label{fig:beambemletreq}}
\end{figure}

In Section II, we briefly discussed the possible use of a properly designed dispersive section to remove the correlation and produce a train of microbunches. In fact, another scheme is to choose the proper initial horizontal C-S parameters upstream of the EEX beamline such that the final longitudinal trace spaces associated to each beamlet are uncorrelated downstream of the EEX beamline. For the sake of simplicity we consider a simple model for the EEX beamline where the deflecting cavity is described by its thin-lens approximation; see Eq.~\ref{eq:thinlenstransform}.  Imposing the final longitudinal trace space associated to, e.g., the central beamlet to be uncorrelated ($\mean{z\delta}=0$) yields a condition between the beamletsÕ horizontal C-S parameters ($\alpha_{s,0}$ and $\beta_{s,0}$) upstream of the exchanger 
\begin{eqnarray} \label{eqn:betaoptimi}
\beta_{s,0}= \frac{L\nu}{2}\alpha_{s,0} \pm \left[\left(\frac{L\nu}{2}\right)^2 - \nu L^2\eta (1+\alpha_{s,0}^2)\right]^{1/2},
\end{eqnarray}
where $\nu\equiv(1-\eta^2)/(L\xi)$. For $\beta_{s,0}$ to be real-positive the condition  $|\alpha_{s,0}|>|1+\nu|/\sqrt{\nu}$ needs to be fulfilled. Equation~\ref{eqn:betaoptimi} represents a hyperbola in the $(\alpha_{s,0},\beta_{s,0})$ space. The horizontal C-S parameters associated to a beamlet generated by intercepting a beam with initial parameters ($\alpha_-,\beta_-$) are $\alpha_{s,-}=0$, $\beta_{s,-}=\beta_-$ and the beamlet's emittance is $\varepsilon_{s,-}=w^2/(12\beta_-)$ where $w$ is the slit width.  Therefore a set of quadrupoles would be needed between the mask and the EEX beamline to match the initial beamlet C-S parameters $(\alpha_-,\beta_-)$ to the required values $(\alpha_{s,0},\beta_{s,0})$ upstream of the EEX beamline. Doing so will also modify the C-S parameters of the entire beam and might result in a significant beam emittance dilution as discussed in Section~\ref{sec:dilutingeffects}. For the geometry considered in this paper (see Tab.~\ref{tab:beamline}) and under the single-particle dynamics it is possible to tune the C-S parameters to get the beamlets' longitudinal trace space upright downstream of the EEX beamline without significantly diluting the emittances; see Fig.~\ref{fig:beambemletreq}. In general, it is always possible to find a beamline configuration upstream of the EEX beamline to simultaneously provide upright trace space for the beamlets while matching the horizontal C-S parameters of the entire beam to values minimizing emittance dilution. This can be done by constructing a beamline that provides the following incoming beam parameter upstream of the slits: 
\begin{eqnarray}
\left[
\begin {array}{c}
\beta_-  \\
\alpha_-
\end {array}
\right]=
\left[
\begin {array}{cc}
\sqrt{\frac{\beta_{s,0}}{\beta_s}}  & 0 \\\noalign{\medskip}
\frac{\alpha_{s,0}-\alpha_s}{\sqrt{\beta_{s,0}\beta_s}}  & \sqrt{\frac{\beta_{s}}{\beta_{s,0}}} \\\noalign{\medskip}
\end {array}
\right]
\left[
\begin {array}{c}
\beta_0  \\
\alpha_0
\end {array}
\right].
\end{eqnarray}
Therefore inserting the mask between two matching sections, each minimally composed of four quadrupoles, enables to control the beamlets longitudinal trace space correlation downstream of the EEX beamline while minimizing beam emittance dilution during transport through the EEX beamline.

\subsection{Arbitrary current profiles\label{sec:arbitraryprofileideal}}
\begin{figure}[hh]
\includegraphics[scale = 0.47]{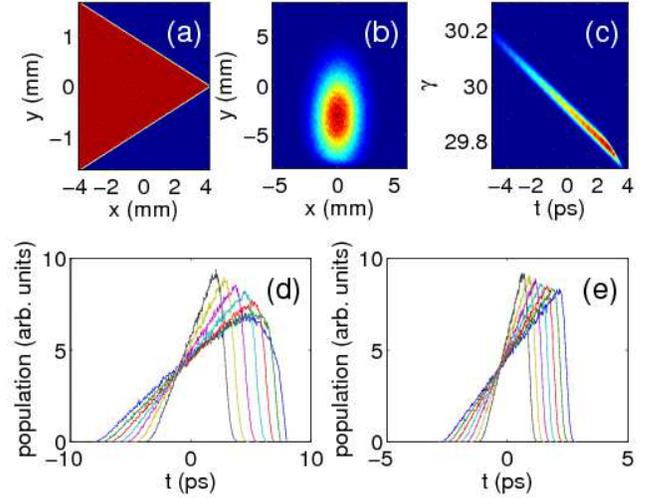}
\caption{Example of the generation of linearly-ramped current profile from an initial uniform triangular distribution in the $(x,y)$ space (a) shown with corresponding final transverse distribution (b) and longitudinal phase space (c) downstream of the EEX beamline. The initial normalized emittance partition is $(\varepsilon^n_{x,0},\varepsilon^n_{z,0})=(10,1)$~$\mu$m and $(\alpha_{x,0}, \beta_{x,0},)=( 2, 12~\mbox{m})$. The current profiles downstream of the EEX beamline are shown in plots (d) and (e) for respectively $(\varepsilon^n_{x,0},\varepsilon^n_{z,0})=(10,1)$ and (1,10)~$\mu$m. The incoming horizontal C-S parameters are a betatron function of 12~m and the $\alpha$-function was varied by increment of 1 from -4 to +2 [corresponding to the seven traces shown in plots (d) and (e)]. The initial beam Lorentz factor for these simulations is $\gamma=30$.  The head of the bunch corresponds to $t<0$. 
\label{fig:ramped}}
\end{figure}

The generation of arbitrarily-shaped current profiles could be achieved using a mask technique as described in the previous section. As an illustrative example, we first consider the academic case of a  transverse density following a uniform triangular distribution; see Fig.~\ref{fig:ramped} (a). Such a distribution results in a linearly-ramped current profile downstream of the EEX beamline upon proper choice of C-S parameter; see Fig.~\ref{fig:ramped} (d,e). The latter figure clearly demonstrates the inability of the pulse shaping technique to resolve the initial high-frequency features (in this case the hard edge associated to the triangular profile). As discussed in Section~\ref{sec:intro} this limitation stems from the final longitudinal emittance values. In the present case, for the same final longitudinal betatron function, the beam with larger final longitudinal emittance has its temporal profile significantly smeared [see Fig.~\ref{fig:ramped}(d)] compared to the beam with lower longitudinal emittance [see Fig.~\ref{fig:ramped}(e)]. 

\begin{figure}[hh]
\includegraphics[scale = 0.47]{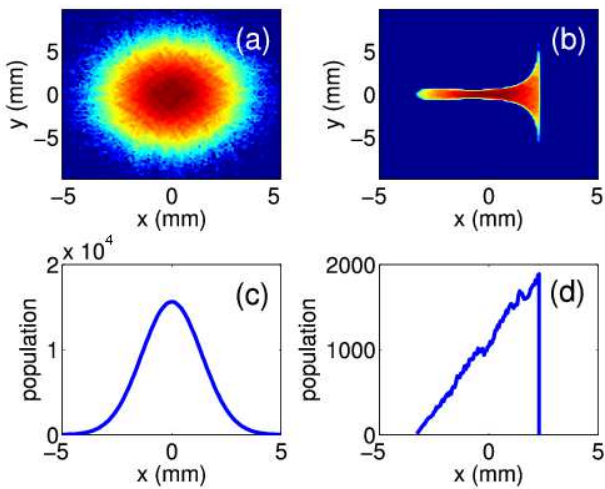}
\caption{Example of mask shape needed to generate a linearly-ramped horizontal profile by interception of beam with transverse radial Gaussian density (a). The mask profile used is described by  Eq.~\ref{eqn:nu} and only macroparticles with $|y| \le \nu(x)$ are kept resulting in the transverse charge density shown in (b). The initial Gaussian horizontal profile (c) is transformed into a linear-ramped horizontal profile (d). For this example the initial transverse beam size are $\sigma_{x}=1.4$~mm and $\sigma_y=2.5$~mm and the parameters  in Eq.~\ref{eqn:nu} are $g_0=100$ and $g_1=3\times 10^4$. \label{fig:maskshape}}
\end{figure}

For realistic beams , a mask with suitable shape needs to be used. Taking the case of a initial Gaussian horizontal profile and imposing the horizontal profile downstream of the mask to be a linear function of $z$ [$P(x) \propto x$] results in the mask shape
\begin{eqnarray}\label{eqn:nu}
\nu(x) = \sqrt{2}\sigma_{y} \mbox{erfinv}\left[-\sigma_x (g_0 + g_1 x) \exp \left(\frac{x^2}{2\sigma_x^2}\right)\right], 
\end{eqnarray}
where $g_0$ and $g_1$ are constants that control the offset and steepness of the linear ramp. The mask profile along with the initial and final horizontal profiles are depicted in Fig.~\ref{fig:maskshape}. For this particular case of an initial Gaussian distribution, only $\sim 19$\% of the incoming charge is transmitted through the mask thus making the masking technique highly inefficient. The case of an initial uniform circular distribution is more favorable but still results in excessive charge loss.  

In principle any current profile can be synthesized  with the proposed technique.  The main disadvantage of using a mask being the substantial charge loss. An alternative method is to shape the beam using the technique discussed later in Section~\ref{sec:rampedawa}. 

\begin{figure*}[t]
 \includegraphics[scale = 0.36]{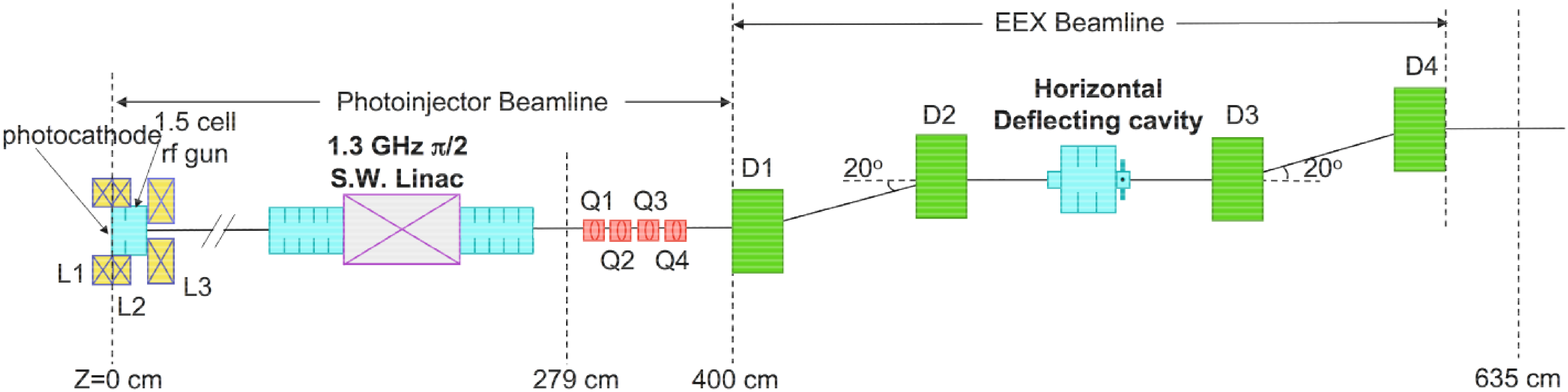}
\caption{Overview of the photoinjector configuration used for realistic simulations of bunch shaping. The legend is ``D$i$" dipole magnets, ``Q$i$" quadrupole magnets,  ``L$i$" solenoidal lenses. The geometry of the EEX beamline is detailed in Table~\ref{tab:beamline}. \label{fig:awa}}
\end{figure*}
\section{Possible Applications in low-energy accelerators\label{sec:appl1}}
In this Section we demonstrate, via start-to-end simulations, possible use of the current shaping technique described in this Paper. We use as a simulation platform the AWA facility shown in Fig.~\ref{fig:awa}.  The accelerator incorporates a photoemission source consisting of a 1+1/2 cell rf cavity operating at $f = 1.3$~GHz, henceforth referred to as rf gun. An ultraviolet (uv) laser beam impinges a magnesium photocathode located on the back plate of the rf gun half cell. The thereby photoemitted electron bunch exits from the rf gun with a maximum kinetic energy of approximately 8~MeV and is injected into a 10-cell standing wave normal conducting booster cavity operating on the $\pi/2$ mode at $f = 1.3$~GHz. The final beam energy can be up to $\sim 16$~MeV. The rf gun is surrounded by three solenoidal magnetic lenses independently powered referred to as L1, L2 and L3; see Fig.~\ref{fig:awa}. \\

The simulations of the EEX beamline presented in this Section were performed with the particle-in-cell (PIC) program {\sc impact-t} which includes space charge effects by solving Poisson's equation in the bunch's rest frame using a three-dimensional quasi-static algorithm~\cite{impact}. The deflecting mode cavity was simulated with {\sc microwave-studio\textregistered} and the generated 3-D electromagnetic field maps were imported  in {\sc impact-t}; see Appendix A. The simulated magnetic field of the horizontally-bending rectangular dipoles was used to  accurately model the dipoles in {\sc impact-t}  by representing the fringe field regions with eight-order Enge functions~\cite{enge}.

\subsection{Terahertz radiation source}
Terahertz (THz) radiation occupies a very large portion of the electromagnetic spectrum and has generated much recent interest due to its ability to penetrate deep into many organic materials without the damage associated with ionizing radiation such as x-rays. THz radiation is easily produced in laser-based method via optical rectification~\cite{Orectification0, Orectification1, Orectification2}. The radiation spectrum is generally broadband and  the associated energy is of the order of nJ per pulse. Accelerator-based THz radiation sources present advantages in term of emitted power, frequency tunability and spectral bandwidth. The spectral fluence radiated by a bunch of $N_e$ electrons taken to follow a line-charge distribution is~\cite{saxon}
\begin{eqnarray}
\left[\frac{dP}{d\omega d\Omega }\right]_{N_e} &=& [\frac{dP}{d\omega d\Omega
}]_{1} \left[N_e + N_e(N_e-1)\widetilde{F}(\omega)\right],
\end{eqnarray}
where $[\frac{dP}{d\omega d\Omega }]_1$ is the single electron power density associated to the considered electromagnetic process and $\widetilde{F}(\omega) \equiv |b(k=\omega/c)|^2$ is the bunch form factor (BFF).  Considering a series of $N_b$ identical microbunches  with normalized distribution $\Lambda(t)$  we have $S(t)=N_b^{-1} \sum_{n=1}^{N_b} \Lambda(t+nT)$ (where $T\equiv\lambda_m/ c$ is the period and $\lambda_m$ the density modulation wavelength defined earlier) giving  $|S(\omega)|^2 =  \Xi  |\Lambda(\omega)|^2$. The intra-bunch coherence factor  $\Xi\equiv N_b^{-2} \sin^2(\omega N_b T/2)/[\sin^2(\omega T/2)]$ describes the enhancement of radiation emission at resonance, i.e. for frequencies $\omega = 2\pi p c/\lambda_m$ (where $p$ is an integer). 

In order to demonstrate the generation of a train of microbunch or modulated bunches using realistic distributions produced by a photoinjector, we modeled the AWA photoinjector from the photocathode up to downstream of the booster cavity with {\sc astra} (up to $z=2.79$~m in Fig.~\ref{fig:awa}), a PIC program that incorporates a cylindrical-symmetric space charge algorithm~\cite{astra}. The generated 1-nC bunch was collimated through a set of vertical slits thereby producing beamlets and its final horizontal C-S parameters upstream of the EEX beamline were matched to specific values using a set of four quadrupoles shown as Q1, Q2, Q3, and Q4 in Fig.~\ref{fig:awa}. The tracking from $z=2.79$~m up to $z\simeq 6.4$~m was performed with {\sc impact-t} and includes space charge effects. The $N$-macroparticles ensemble downstream of the EEX beamline was represented as a Klimontovich distribution 
\begin{eqnarray} \label{eqn:klimontovich}
P(z)=\frac{1}{N}\sum_{j=1}^N \delta(z-z_j)
\end{eqnarray} 
along the longitudinal coordinate.  The associated BFF is calculated as the summation
\begin{eqnarray}
\widetilde{F}({\omega})=\bigg| \sum_{j=1}^N \sin({\omega}{z}_j/c) \bigg|^2 + \bigg| \sum_{j=1}^N \cos({\omega}{z}_j/c) \bigg|^2. 
\end{eqnarray}

\begin{figure}[hhhhhhhhh!!!!!!!!!!!!!!]
\includegraphics[scale = 0.43]{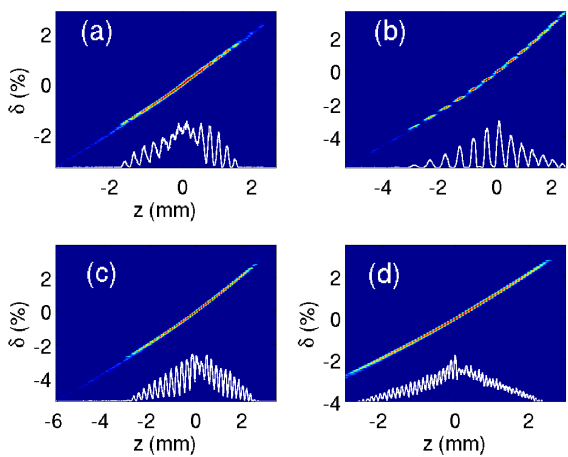}
\caption{Example of conversion of modulated longitudinal trace space downstream of the EEX beamline. The four plots corresponds to the four cases of initial parameters ($\delta w,w,{\cal F}$) of (100, 300, 0.92) [plot (a)], (100, 300, 1.25) [plot (b)], (50, 150, 0.75) [plot (c)] and (25, 75, 0.75) [plot (d)] where $\delta w$ and $w$ are respectively the slit edge-to-edge separation and width in micrometers, and ${\cal F}$ is a scaling factor for the quadrupoles Q1, Q2, Q3, and Q4 (when ${\cal F}=1$ the quadrupoles are tuned to match the horizontal C-S parameter to $(\alpha_{x,0}, \beta_{x,0})=(2,12~\mbox{m})$ upstream of the exchanger). 
\label{fig:train}}
\end{figure}

\begin{figure}[hhhhhhhhh!!!!!!!!!!!!!!]
\includegraphics[scale = 0.48]{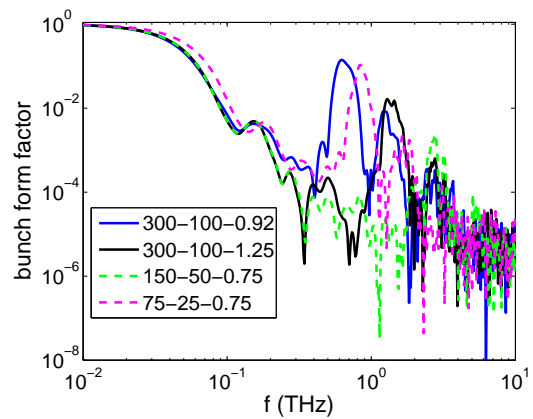}
\caption{Bunch form factor associated to train of microbunches obtained downstream of the EEX beamline. The four traces correspond to four settings/configuration indicated in the legend as ``$\delta w-w-{\cal F}$" where $\delta w$ and $w$ are respectively the slit edge-to-edge separation and width in micrometers, and ${\cal F}$ is a scaling factor for the quadrupoles Q1, Q2, Q3, and Q4 (when ${\cal F}=1$ the quadrupoles are tuned to match the horizontal C-S parameter to $(\alpha_{x,0}, \beta_{x,0})=(2,12~\mbox{m})$ upstream of the exchanger).
\label{fig:bff}}
\end{figure}

Figure~\ref{fig:train} depicts examples of  modulated bunches (the charge is 300~pC downstream of the slits). The BFF associated are enhanced at the fundamental bunching and harmonic frequencies which can be varied by either changing the quadrupoles settings upstream of the EEX beamline or using different slits separations; see Fig.\ref{fig:bff}. In this particular example the frequency can in principle be continuously varied from $\sim 0.6$ to $\sim 2.2$~THz (if we consider the fundamental bunching frequency only; higher frequencies can be obtained by proper filtering to select harmonics of the bunching frequency). We should note that the calculated BFF is representative of the radiation spectrum as long as the line charge distribution assumption is fulfilled, e.g. provided the longitudinal and transverse $\sigma_{\perp}$ rms bunch sizes satisfy $\sigma_z \gg \sigma_{\perp}/\gamma$ if the considered radiation process is transition radiation~\cite{tr}. The generated train of microbunches can also be used to produce super-radiant undulator radiation~\cite{gover}. 

\subsection{Ramped drive beams and bunch trains for beam-driven wakefield accelerators \label{sec:rampedawa}}
Another application of pulse shaping we consider is the generation of beams with linearly-ramped current profiles. Ramped beams maximize the transformer ratio in collinear beam-driven acceleration schemes. A practical implementation using an intercepting mask seems highly inefficient as discussed in Section~\ref{sec:arbitraryprofileideal}. Instead we explore the generation of such a beam by generating a beam with a suitable transverse distribution starting at the photocathode. Several types of distribution were considered for the initial laser transverse distribution and we found that a distribution with a dipole moment in provides the desired ramped horizontal profile upstream of the EEX beamline. We therefore consider laser transverse densities on the cathode described by~\cite{marwanpac07}
\begin{eqnarray}
\rho(r,\theta) = \rho_0[1+ {\cal A} \frac{r}{r_c} \cos (\theta -\theta_0)]
\end{eqnarray}
where $\rho_0(r,\theta)=1$ over a specified domain and $0$ elsewhere, $r_c$ is the beam size on the photocathode, $\theta_0$ is the initial rotation angle of the dipole pattern, and ${\cal A}$ is the amplitude of the perturbation.  The symmetry axis of the laser distribution is rotated such that after propagation through the emittance-compensating solenoid the $x-y$ coupling is removed. This is accomplished by choosing the rotation angle to be~\cite{reiser2}:
\begin{eqnarray}
\theta_0 = \int_0^{\infty}[{eB(r=0,z)}]/[{2m\gamma(z)\beta(z) c}]dz
\end{eqnarray}
where $B(r=0,z)$ is the axial component of magnetic field experienced by the beam, $\gamma$ the beam's Lorentz factor and $\beta \equiv
[1- \gamma^{-2}]^{1/2}$.  

\begin{figure}[hhhhhhhhh!!!!!!!!!!!!!!]
\includegraphics[scale = 0.48]{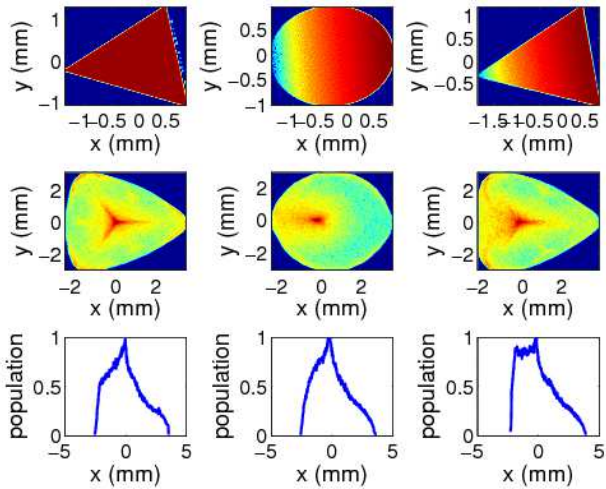}
\caption{Initial transverse densities on the photocathode surface (top row) and corresponding transverse densities (middle row) and horizontal profiles (bottom row) at $z=2.79$~m. The three column respectively correspond to the case of a uniform-triangle distribution (left column), a circular distribution with a dipole moment (middle column) and a triangular-shaped distribution with a dipole moment (right column).  \label{fig:diploleevolution}}
\end{figure}

Three types of distributions were considered: (1) an initial uniform triangle ($\rho_0=1$ inside a equilateral triangle and ${\cal A}=0$), (2) a circular distribution with a dipole moment ($\rho_0=1$ for $r<r_c$ and $\forall \theta$ and ${\cal A}=1$), and (3) a triangular-shaped distribution with a dipole moment (${\cal A}=1$). For all cases the distribution on the cathode was centered w.r.t to its barycenter and the rms transverse sizes were forced to be 1~mm. These three types of distribution can be implemented using uv masks of variable transmission~\cite{dowell}. The laser density on the photocathode surface and the resulting 1-nC bunch at $z=2.79$~m are shown in Fig.~\ref{fig:diploleevolution}. 

\begin{figure}[hhhhhhhhh!!!!!!!!!!!!!!]
\includegraphics[scale = 0.48]{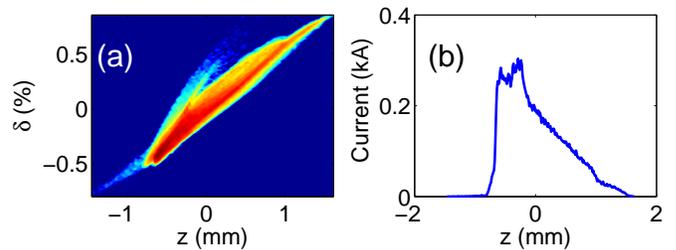}
\caption{Example of longitudinal phase space (left) and associated linearly-ramped current profile (b) obtained downstream of the EEX beamline. The beam Lorentz factor is $\gamma\simeq 33.4$  and the rms bunch length is $490$~$\mu$m. The initial distribution used for this calculation is the triangular-shaped distribution with a dipole moment (right column in Fig.~\ref{fig:diploleevolution}).~\label{fig:dipoleEEX}}
\end{figure}

Figure~\ref{fig:diploleevolution} demonstrates that these distributions have a ramped horizontal profile upstream of the EEX beamline. For the considered accelerator settings, the triangular-shaped distribution with a dipole moment yields the best  linearly-ramped horizontal profile that can be converted into a linearly-ramped current profile downstream of the EEX beamline as illustrated for a 1-nC bunch in Fig.~\ref{fig:dipoleEEX}. The quadrupoles Q1, Q2, Q3, and Q4 were tuned to approximately provide C-S horizontal parameters upstream of the EEX beamline close to $(2, 12~\mbox{m})$ (the matching was done in the single-particle dynamics mode and the corresponding quadrupole settings were used in {\sc impact-t} when tracking with space charge). 

\begin{figure}[hhhhhhhhh!!!!!!!!!!!!!!]
\includegraphics[scale = 0.45]{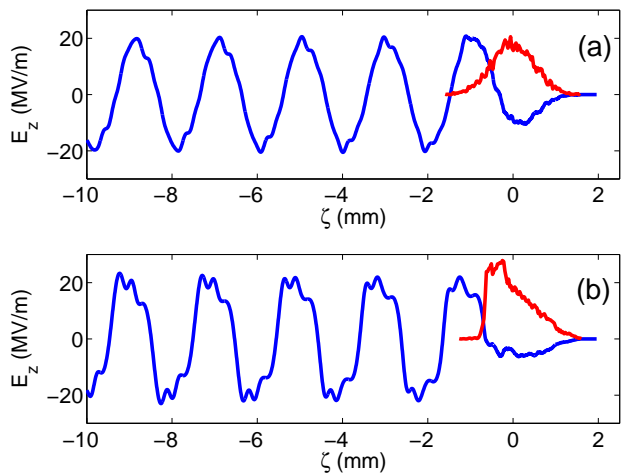}
\caption{Axial electric field (blue traces) behind a 1-nC bunch with Gaussian (a) and linearly-ramped (b) current profile (red traces). In our convention $E_z<0$ corresponds to a decelerating field. For both case the bunch is centered at $\zeta=0$ and the head corresponds to $\zeta>0$. The transformer ratio quoted in the text is computed as the ratio of $E_z(\zeta\simeq -1.3~\mbox{mm})$ over $|E_z(\zeta\simeq 0~\mbox{mm})|$.~\label{fig:wak}}
\end{figure}

The final normalized emittance partition $(\varepsilon^n_x,\varepsilon^n_y,\varepsilon^n_z)=(43.6, 5.8,12.1)$~$\mu$m makes such a bunch suitable as a drive beam for beam-driven techniques based on asymmetric structures, e.g., dielectric slabs~\cite{tremaine,xiao}. The final current distribution is used to estimate the achievable axial wakefield following Ref.~\cite{tremaine}. We consider a slab structure with half gap $a=0.150$~mm and dielectric thickness $b-a=0.150$~mm We take the relative dielectric permittivity to be $\epsilon=9.8$ and compute the axial wakefield produced by the bunch as 
\begin{eqnarray}
E_z(\zeta)=\sum_{i-1}^n \sum_{j=1}^N {w}_i (\zeta -\zeta_j), 
\end{eqnarray}
where $\zeta\equiv z-v_b t$ ($v_b\simeq c$ is the bunch's velocity), $w_i(\zeta)$ is the Green's function for the $i$-th mode wakefield described by Eq.~2.6 in Ref.~\cite{tremaine}. The indices $i$ and $j$ in the latter equation correspond respectively to the summations over the number of modes considered (we take $n=10$) and the number of macroparticles ($N=2\times 10^5$) representing the bunch (here we model  the bunch as a Klimontovich distribution; see Eq.~\ref{eqn:klimontovich}). The computed axial electric field is shown and compared with the electric field induced by a Gaussian beam with similar rms duration in Fig.~\ref{fig:wak} . The transformer ratio defined as the ratio of the maximum accelerating field to the maximum (in absolute value) decelerating field is found to be ${\cal R} \simeq 20.08/10.22 = 1.96$ and ${\cal R} \simeq 22.05/6.13 = 3.60$ for respectively the Gaussian and linearly-ramped bunch. 

If a round beam is desired downstream of the EEX beamline, the shaping technique would need to be combined with the flat beam technique~\cite{brinkmann,piotflat}.

Finally, a bunch with linearly-ramped horizontal profiles could also be intercepted with a series of slits to generate a train of bunch. This would result in a ramped bunch train and has application to resonantly excite wakefield in beam-driven acceleration schemes.  

%\section{Possible applications and configuration in high-energy accelerators\label{sec:appl2}}
%
%The technique described in this paper could in principle be used to shape a high-energy beam or %produced train of microbunch as recently discussed in Ref.~\cite{kip}. One of the inherent problems comes from the asymmetric final transverse emittance values (since the longitudinal emittance is mapped onto the horizontal emittance). In this Section we briefly discuss a configuration consisting of back-to-back two EEX beamlines as an alternative solution; see Fig.~\ref{doubleEEX}. In this settings the beam is first transformed such that its final horizontal profile corresponds to the initial longitudinal profile and shaped. A second EEX beamline converts back the shaped transverse profile into a shaped current profile. 
%
\section{summary}
We presented a general method for generating relativistic electron beams with arbitrary current profiles. The proposed method offers a very generic tool for precisely controlling the current distribution (and possibly the emittance partitions) to match the requirements imposed by the front end application. The technique takes advantage of the recently proposed transverse-to-longitudinal phase space exchange to map the horizontal profile associated to the electron bunch into its current profile. In principle any current profile can be produced by properly manipulating the bunch's transverse density to provide the desired horizontal profile  upstream of the EEX beamline. We have considered special cases and showed how the method could be used to produce a train of microbunches with variable spacing and linearly-ramped bunch using start-to-end simulations of a realistic configuration in a $\sim 16$~MeV photoinjector. The technique could also be applied to high-energy beams as discussed, for instance, in Ref.~\cite{kip}. At high energies, the use of a thick mask to intercept the beam might be prohibited and instead a selection technique based on emittance spoiling using a thin foil might be preferable~\cite{emmaspoiled}. 

{\em Note added in proof} $-$ Very recently an experiment conducted at the Fermilab's A0 photoinjector demonstrated some of the possibilities discussed in this paper~\cite{yinesunmubunch} including the generation of a train of sub-picosecond bunches with variable separation~\cite{yineprl}. 

\section{acknowledgments}
We are thankful to the members of the ANL-FNAL-NIU emittance exchange collaboration, in particular to M. Church, H. T. Edward, W. Gai and K.-J. Kim, for encouragement and fruitful discussions. We thank  C. Prokop (NIU) for his comments on the manuscript. The work of M.R. and P.P. was supported by the US Department of Energy under Contract No. DE-FG02-08ER41532  with Northern Illinois University.  The work of Y.-E S. and partially of P.P is supported by the Fermi Research Alliance, LLC under Contract No. DE-AC02-07CH11359 with the U.S. Department of Energy. The work of J. G. P was supported by Argonne National Laboratory under Contract No. DE-AC02-06CH11357 with the U.S. Department of Energy. 

\section*{APPENDIX A: NUMERICAL VALIDATION OF THE ANALYTICAL MATRIX OF A DEFLECTING CAVITY }
In this appendix we show that  the transfer matrix of a deflecting cavity derived in Ref.\cite{edwardsNote} for an ideal pillbox cavity operating on the $TM_{110}$ accurately describes the matrix
modeled numerically computed using the electromagnetic field obtained from {\sc microwave-studio\textregistered}  simulations.

\begin{figure}[hhhhhh]
 \includegraphics[scale = 0.33]{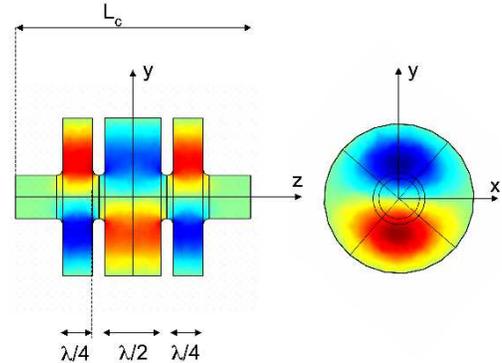}
\caption{Geometry of the TM$_{010}$-like deflecting mode cavity used for the simulation presented in this paper. The superimposed false-color pattern represents the value of the $E_z$ field in the ($x=0, y,z$) plane (the blue and red colors respectively correspond to the maximum and minimum values of the field). \label{fig:cavity}}
\end{figure}

We consider a cylindrical-symmetric pillbox cavity operating on the $TM_{110}$ at the zero-crossing phase. The corresponding electromagnetic field components in the paraxial approximation are given by
\begin{eqnarray}
\begin{array}{ccc}
E_z(x,y,z,t) &=& {E'_0 x} \cos(\omega t)  \\
B_y(x,y,z,t) &=& \frac{E'_0}{\omega} \sin (\omega t), \\
\end{array}
\end{eqnarray}
where $E_0'\equiv \partial E_x /\partial x \simeq (2\pi/\lambda) E_0$ (where $E_0$ is the peak electric field) and $\omega\equiv2\pi f$ ($f$ is the operating frequency).  The thick-lens transfer matrix in $(x,x',z,\delta)$ is given by Eq.~\ref{eq:cavdon} where $\kappa\equiv \frac{2\pi}{\lambda}\frac{ eE_0}{pc}$. The cavity used in our simulations is shown in Fig.~\ref{fig:cavity}. It operates at $f=1.3$~GHz and  consists of three cells of length $(1/2-1-1/2)\times \lambda/2$. Two end pipes were added in the electromagnetic model to properly account for the fringe fields of the cavity.

Considering the cavity to consists of three uncoupled pillbox resonators of length $\lambda/4$, $\lambda/2$ and $\lambda/4$ we compute the transfer matrix to be
\begin{eqnarray}\label{eq:cav}
R \equiv
\left[ \begin {array}{cccc}
R_{11} & R_{12} & R_{15} & R_{16} \\
R_{21} & R_{22} & R_{25} & R_{26} \\
R_{51} & R_{52} & R_{55} & R_{56} \\
R_{61} & R_{62} & R_{65} & R_{66} \\
\end {array} \right] =
\left[ \begin {array}{cccc}
1 & L_c & \frac{\kappa L_c}{2} & 0 \\
0 &  1   & \kappa & 0  \\
0 &  0   &  1  & 0   \\
\kappa &  \frac{\kappa L_c}{2}  &  \frac{23 \kappa^2 \lambda}{128}  &  1
\end {array} \right], 
\end{eqnarray}
where $L_c$ is the cavity length including the outer pipes; see Fig.~\ref{fig:cavity}.

To numerically evaluate  the first order transfer matrix of the cavity, we track seven macroparticles in the cavity showed in Fig.~\ref{fig:cavity}. Six of the macroparticles are displaced by an infinitesimal small amount  ${\bf \delta X}^i$ with respect to the reference macroparticle along one of the phase space coordinates. This initial displacement is mapped  downstream of the cavity  as a displacement ${{\bf \delta X'}^j}$ with respect to the final position of the reference macroparticle. The initial and final displacement are related via
\begin{eqnarray}
{{\bf \delta X'}_j} = \sum_{i=1}^6 R_{ij} {\bf \delta X}_i. 
\end{eqnarray}
Inversion of the latter system of equations directly provides the transfer matrix elements.

\begin{figure}[hhhhhhhhh]
 \includegraphics[scale = 0.48]{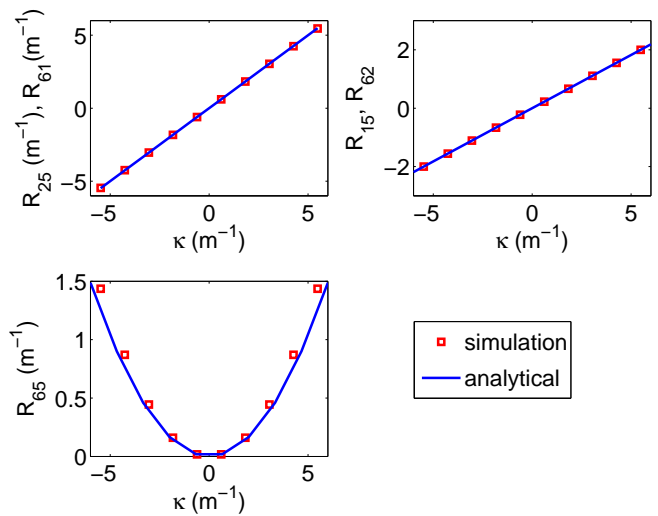}
\caption{Comparison of the analytical (solid lines) and numerical (squares) calculations of the $\kappa$-dependent elements of transfer matrix associated to the deflecting cavity considered in this paper.  \label{fig:cavmatrix}}
\end{figure}

The $\kappa$-dependent elements of the numerically-computed transfer matrix are compared with those analytically calculated using Eq.~\ref{eq:cav} in Fig.~\ref{fig:cavmatrix}. The agreement between the numerical and analytical models is excellent: the relative discrepancy is $<5$\% over $\kappa \in [-5, 5]$~m$^{-1}$.
%

%\newpage


\begin{thebibliography}{99}
\bibitem{CSTMW}  {\sc microwave studio} is a 3D full-wave electromagnetic field simulation commercially available from Computer Simulation Technologies AG.
\end{thebibliography}
\end{document}